\documentclass[twocolumn,10pt]{article}
\usepackage[cmex10]{amsmath}
\usepackage{amsthm}
\usepackage{amsfonts}
\usepackage{geometry}
\usepackage{graphicx}
\usepackage{xcolor}
\usepackage{url}
\usepackage{mdwlist}
\usepackage[noend]{algorithmic}
\usepackage[normalem]{ulem}

\title{A Secure Submission System for Online Whistleblowing Platforms}

\author{%
  Volker Roth, Benjamin G\"uldenring, Eleanor Rieffel,$^{\dag}$ Sven
  Dietrich,$^{\ddag}$ Lars Ries\\[1.2ex]
  Freie Universit\"at Berlin, $^\dag$FX Palo Alto Laboratory,\\
  $^\ddag$Stevens Institute of Technology%
}
\date{\today}%
\DeclareMathOperator{\Enc}{Enc}

\DeclareMathOperator{\EncData}{EncData}
\DeclareMathOperator{\EncZero}{EncZero}
\DeclareMathOperator{\DecVrfy}{DecVrfy}

\DeclareMathOperator{\imod}{mod}

\newcommand{\Zm}[1][N]{\mathbb{Z}^{\ast}_#1}
\newcommand{\Za}[1][N^s]{\mathbb{Z}_{#1}}
\newcommand{\Zn}[1][N^{s+1}]{\mathbb{Z}^{\ast}_{#1}}

\newcommand{\Ktestmean}{0.011578}
\newcommand{\Kdecmean}{0.192843}

\newcommand{\Keccmean}{0.006632}
\newcommand{\Ueccmean}{ms}
\newcommand{\Keccsdev}{0.000171}
\newcommand{\Ueccsdev}{ms}
\newcommand{\Keccklen}{256}

\newcommand{\Kodjlmean}{7.55}

\newcommand{\Kodjmmean}{14.67}

\newcommand{\Kodjhmean}{28.68}

\newcommand{\Kodjlsdev}{0.06}

\newcommand{\Kodjmsdev}{0.05}

\newcommand{\Kodjhsdev}{0.38}

\newcommand{\Kads}{50}

\newcommand{\Kp}{0.9}
\newcommand{\Kfile}{2}
\newcommand{\Ufile}{MB}
\newcommand{\Kheaders}{400}
\newcommand{\Ktlen}{4496}
\newcommand{\Kloadinc}{220}
\newcommand{\Uloadinc}{KB/user}
\newcommand{\Kclen}{3072}
\newcommand{\Uclen}{bytes}
\newcommand{\Kmlen}{2303}
\newcommand{\Umlen}{bytes}
\newcommand{\Kcblk}{911}

\newcommand{\Kcnum}{1010}

\newcommand{\Kdays}{21}

\newcommand{\Kusers}{138}
\newcommand{\Uusers}{million}
\newcommand{\Khours}{11}

\newcommand{\Kmaxreqi}{8000}
\newcommand{\Umaxreqi}{reqs/s}
\newcommand{\Kmaxreqii}{8500}

\newcommand{\Keccreqii}{8046}
\newcommand{\Ueccreqii}{reqs/s}
\newcommand{\Kmbsii}{209}
\newcommand{\Umbsii}{Mb/s}
\newcommand{\Kreqi}{174243}
\newcommand{\Ureqi}{reqs/s}
\newcommand{\Ksreqi}{175495}
\newcommand{\Usreqi}{reqs/s}
\newcommand{\Kunitsi}{22}

\newcommand{\Kunitsii}{22}

\newcommand{\Kcapiii}{18}
\newcommand{\Ucapiii}{Mb/s}
\newcommand{\Kpktiii}{768}

\newcommand{\Kmaxdata}{82}
\newcommand{\Umaxdata}{reqs/s}
\newcommand{\Kavgdata}{65}
\newcommand{\Uavgdata}{reqs/s}

\newcommand{\Kcores}{11}
\newcommand{\Kpdata}{0.11}

\newcommand{\Ksavdec}{0.61}

\newcommand{\Kunitsiii}{2}
\newcommand{\Kmaxwb}{51480}
\newcommand{\Kunitwb}{25740}
\newcommand{\Kmaxfiles}{2827}

\newcommand{\Kcpm}{0.25}
\newcommand{\Ucpm}{USD}
\newcommand{\Klowads}{5}

\newcommand{\Kunitmonth}{400}
\newcommand{\Uunitmonth}{USD/month}

\newcommand{\Kdailyusd}{579}
\newcommand{\Udailyusd}{USD/day}

\newcommand{\Kbreakeven}{0.34}
\newcommand{\Ubreakeven}{percent}

\begin{document}
\maketitle

\begin{abstract}
  Whistleblower laws protect individuals who inform the public or an
  authority about governmental or corporate misconduct.  Despite these laws,
  whistleblowers frequently risk reprisals and sites such as WikiLeaks
  emerged to provide a level of anonymity to these individuals.  However, as
  countries increase their level of network surveillance and Internet
  protocol data retention, the mere act of using anonymizing software such
  as Tor, or accessing a whistleblowing website through an SSL channel might
  be incriminating enough to lead to investigations and repercussions.  As
  an alternative submission system we propose an online advertising network
  called \emph{AdLeaks.}  AdLeaks leverages the ubiquity of unsolicited
  online advertising to provide complete sender unobservability when
  submitting disclosures.  AdLeaks ads compute a random function in a
  browser and submit the outcome to the AdLeaks infrastructure.  Such a
  whistleblower's browser replaces the output with encrypted information so
  that the transmission is indistinguishable from that of a regular browser.
  Its back-end design assures that AdLeaks must process only a fraction of
  the resulting traffic in order to receive disclosures with high
  probability.  We describe the design of AdLeaks and evaluate its
  performance through analysis and experimentation.
\end{abstract}

\section{Introduction}


Corporate or official corruption and malfeasance can be difficult to uncover
without information provided by insiders, so-called \emph{whistleblowers.}
Even though many countries have enacted, or intend to enact, laws meant to
make it safe for whistleblowers to disclose
misconduct~\cite{Banisar2009,OsterhausF2009}, whistleblowers fear
discrimination and retaliatory action regardless, and sometimes justifiably
so~\cite{ERC2012,Lennane1993}.

It is therefore unsurprising that whistleblowers often prefer to blow the
whistle anonymously through other channels than those mandated by
whistleblowing legislature.  This gave rise to whistleblowing websites such
as \emph{Wikileaks.}  However, the proliferation of surveillance technology
and the retention of Internet protocol data records~\cite{BertholdBK2009}
has a chilling effect on potential whistleblowers.  The mere act of
connecting to a pertinent Website may suffice to raise suspicion
\cite{Gustin2010}, leading to cautionary advice for potential
whistleblowers.

The current best practice for online submissions is to use an
SSL~\cite{rfc6101} connection over an anonymizing network such as
Tor~\cite{DingledineMS2004}.  This hides the end points of the connection
and it protects against malicious exit nodes and Internet Service Providers
(ISPs) who may otherwise eavesdrop on or tamper with the connection.
However, this does not protect against an adversary who can see most of the
traffic in a network~\cite{ChakravartySK2010,DyerCRS2012}, such as national
intelligence agencies with a global reach and view.

In this paper, we suggest a submission system for online whistleblowing
platforms that we call \emph{AdLeaks.}  The objective of AdLeaks is to make
whistleblower submissions unobservable even if the adversary sees the entire
network traffic.  A crucial aspect of the AdLeaks design is that it
eliminates any signal of intent that could be interpreted as the desire to
contact an online whistleblowing platform.
AdLeaks is essentially an online advertising network, except that ads carry
additional code that encrypts a zero probabilistically with the AdLeaks
public key and sends the ciphertext back to AdLeaks.  A whistleblower's
browser substitutes the ciphertext with encrypted parts of a disclosure.
The protocol ensures that an adversary who can eavesdrop on the network
communication cannot distinguish between the transmissions of regular
browsers and those of whistleblowers' browsers.  Ads are digitally signed so
that a whistleblower's browser can tell them apart from maliciously injected
code.  Since ads are ubiquitous and there is no opt-in, whistleblowers never
have to navigate to a particular site to communicate with AdLeaks and they
remain unobservable.  Nodes in the AdLeaks network reduce the resulting
traffic by means of an \emph{aggregation} process.  We designed the
aggregation scheme so that a small number of trusted nodes with access to
the decryption keys can recover whistleblowers' submissions with high
probability from the aggregated traffic.  Since neither transmissions nor
the network structure of AdLeaks bear information on who a whistleblower is,
the AdLeaks submission system is immune to passive adversaries who have a
complete view of the network.

In what follows, we detail our threat model and our assumptions, we give an
overview over the design of AdLeaks, we analyze its scalability in detail,
we report on the current state of its implementation and we explain how
AdLeaks uses cryptographic algorithms to achieve its security objectives.

\section{Threat Model}

The primary security objective of AdLeaks is to conceal the \emph{presence}
of whistleblowers, and to eliminate network traces that may make one suspect
more likely than another in a search for a whistleblower.  This security
objective is more important to AdLeaks than, for example, availability.  We
rather risk that a disclosure does not come through than compromise
information about a whistleblower.  In what follows we detail the threats
our system architecture addresses as well as the threats it does not
address.

\subsection{Threats in Our Scope}

AdLeaks addresses the threat of an adversary who has a global view of the
network and the capacity to store or obtain Internet protocol data records
for most communications.  The adversary may even require anonymity services
to retain connection detail records for some time and to provide them on
request.  The adversary may additionally store selected Internet traffic and
he may attempt to mark or modify communicated data.  However, we assume that
the adversary has no control over users' end hosts, and he does not block
Internet traffic or seizes computer equipment without a court order.  We
assume that the court does not \emph{per se} consider organizations that
relay secrets between whistleblowers and journalists as criminal.  The
objective of the adversary is to uncover the identities of whistleblowers.
The threat model we portrayed is an extension of~\cite{BertholdBK2009} and
it is likely already a reality in many modern states, or it is about to
become a reality.  For reasons we explain in the following section we do not
consider additional threats that we would doubtless encounter, for example,
in technologically advanced totalitarian countries.

\subsection{Threats Not in Our Scope}

We exclude blocking from our threat model and our discussion because we do
not contribute to blocking resistance and its inclusion would distract from
our contribution.
The second threat we exclude is that of a flooding attack on our submission
system.  While we have thoughts on how to limit some attacks of this kind we
prefer to make a solid first step towards unobservability before considering
the next step in our research.  We hope that the next step will not become
necessary because this means that countries we believe liberal have gone too
far down the slope towards totalitarianism already.
The third threat we exclude is denial of service by means of fake
transmissions.  This threat manifests at the level of the editorial process
that separates the chaff from the wheat among the potentially many submitted
disclosures.  We consider this threat out of scope in this part of our work.
The fourth threat we exclude is that of malware and spyware.  For example,
sensitive documents in PDF format may contain JavaScript that emits a beacon
whenever the document is viewed.  A careless whistleblower who opens a
sensitive document on his home machine may expose himself or herself in that
fashion~\cite{BowenHKS2009}.  Similarly, if a whistleblower's computer is
infected by a malware or spyware then the whistleblower has no security.

\subsection{Security-Related Assumptions}

AdLeaks ads require a source of randomness in the browser that is suitable
for cryptographic use.  Moreover, the source must be equally good on regular
browsers and on the browsers of whistleblowers.  If a whistleblower's
browser looks decidedly more random than other browsers then whistleblowers
can be readily identified.  Unfortunately the random numbers most browsers
generate are far from random.  However, there are good reasons for browser
developers to support cryptographically secure random number generators in
the near-term~\cite{webcryptoapi}, for example, to prevent illicit user
tracking~\cite{Klein2008}.  In the meantime, entropy collected in the
browser may be folded into a pseudorandom generator using SJCL
\cite{StarkHB09}.  We therefore decided to move forward with our research
assuming that browsers will soon be ready for it.

We assume that whistleblowers use AdLeaks only on private machines to which
employers have no access.  In fact, sending information from work computers
even using work-related e-mail accounts is a mistake whistleblowers make
frequently.  We hope that the software distribution channels we discuss in
Section~\ref{sec:getit} will help reminding whistleblowers to not make that
mistake.

\begin{figure*}[t]
  \centering
  \includegraphics[width=0.8\textwidth]{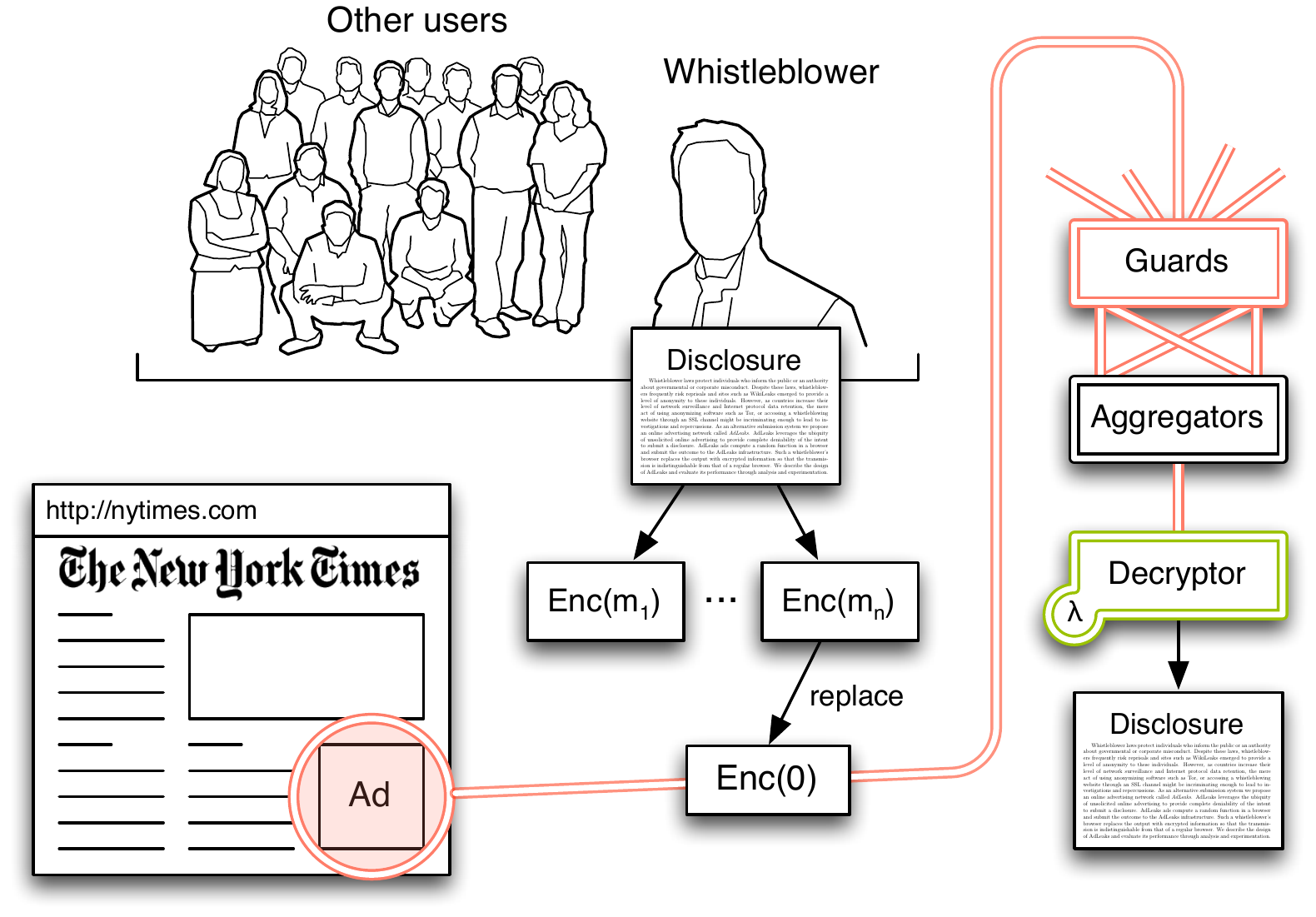}
  \caption{Illustrates the architecture of the AdLeaks system.  Aggregators
    reduce the incoming traffic so that submissions can be funneled to a
    trusted decryptor through a household DSL line.  $\lambda$ denotes the
    decryption key.}
  \label{fig:arch}
\end{figure*}

\section{System Architecture}

AdLeaks consists of two major components.  The first component is an online
advertising network comparable to existing ones.  The network has
advertising partners (the publishers) who include links or scripts in their
web pages which request ads from the AdLeaks network and display them.
Publishers may receive compensation in accordance with common advertising
models, for example, per mille impressions, per click or per lead generated.
Advertisers run campaigns through the AdLeaks network.  AdLeaks may
additionally run campaigns through other ad networks to extend its reach,
for example, funded by donations or profits from its own operations.  The
ecosystem of partners and supporters may include large newspapers, bloggers,
human rights organizations and their affiliates.  For example, Wikileaks has
partnered with organizations such as Der Spiegel, El Pa\'\i{}s and the New
York Times, and OpenLeaks had hinted at support by Greenpeace and other
organizations.  The key ingredient of an AdLeaks ad is not its visual
display but its active JavaScript content.  Supporters who would forfeit
significant revenue when allocating advertising space to AdLeaks ads have a
choice to only embed the JavaScript portion.  The JavaScript is digitally
signed by AdLeaks and contains public encryption keys.

The second major component of AdLeaks is its submission infrastructure.
This infrastructure consists of three tiers of servers.  We refer to these
tiers as \emph{guards, aggregators} and \emph{decryptors.}  When a browser
loads an AdLeaks ad, the embedded JavaScript encrypts a zero
probabilistically with the embedded public key and submits the ciphertext to
a guard.  The guard strips unnecessary encoding and protocol meta-data from
the request and forwards the ciphertext to an aggregator.  An aggregator
aggregates the ciphertexts it receives per second and transmits them to the
decryptor.  What makes this setting challenging is that we want to limit the
bandwidth of the decryptor to a household Internet connection so that we can
keep a close eye on the all-important machine with the decryption keys.  The
aggregation leverages the homomorphic properties of the Damg\aa{}rd-Jurik
(DJ) encryption scheme~\cite{DamgardJN2010}, which means that the product of
the ciphertexts is an encryption of the sum of the plaintexts.  We chose the
DJ scheme because it has a favorable plaintext to ciphertext ratio.

The decryptor decrypts the downloaded ciphertexts and, if it finds data in
them, reassembles the data into files. The files come from whistleblowers.
In order to submit a file, a whistleblower must first \emph{obtain} an
installer that is digitally signed and distributed by AdLeaks.  This is
already a sensitive process that signals intent.  We defer the discussion of
safe distribution channels for the installer to section~\ref{sec:getit}.
\emph{Installing} the obtained software likewise signals the intent to
disclose a secret, and therefore it is crucial that the whistleblower
verifies the signature \emph{before} running the installer, and assures
himself that the signer is indeed AdLeaks.  Otherwise, he is vulnerable to
Trojan Horse software designed to implicate whistleblowers.  When run, the
installer produces an instrumented browser and an encryption tool.  The
whistleblower prepares a file for submission by running the encryption tool
on it.  The tool's output is a sequence of $\ell$ ciphertexts.  Henceforth,
whenever an instrumented browser runs an ad signed by AdLeaks, it replaces
the script's ciphertext with one of the $\ell$ ciphertexts it has not
already used as a replacement.

In order to distinguish ciphertexts that are encryptions of zeros from
ciphertexts that are encryptions of data we refer to the former as
\emph{white} and to the latter as \emph{gray.}  If the aggregator aggregates
a set of white ciphertexts then the outcome is another white one.  If
exactly one gray ciphertext is aggregated with only white ones then the
outcome is gray as well.  If we decrypt the outcome then we either recover
the data or we determine that there was no data to begin with.  If two or
more gray ciphertexts are aggregated then we cannot recover the original
data from the decryption.  We call this event a \emph{collision} and we
refer to such an outcome as a \emph{black} one.  Obviously, we must expect
and cope with collisions in our system.  In what follows, we elaborate on
details of the design that are necessary to turn the general idea into a
feasible and scalable system.

\subsection{Disclosure Preparation}
\label{sec:prep}

In order to handle collisions, the encryption tool breaks a file into blocks
of a fixed equal size and encodes them with a loss tolerant \emph{Fountain
  Code}.  Fountain codes encode $n$ packets into an infinite sequence of
output packets of the same size such that the original packets can be
recovered from any $n'$ of them where $n'$ is only slightly larger than $n$.
For example, a random linear Fountain Code decodes the original packets with
probability $1-\delta$ from about $n+\log_2(1/\delta)$ output
packets~\cite{MacKay2005}.  Let $n''$ be somewhat larger than $n'$ and let
$m_1,\dots,m_{n''}$ be the Fountain encoding of the file.  The tool then
generates a random file identification number $k$ and computes: $c_i
=\Enc_{\kappa_1}^{\text{cca}}(\EncData_{\kappa_2}(m_i, k || i || n))$ for
$1\le i\le n''$ where $\kappa_1$ is an aggregator key and $\kappa_2$ is the
actual submission key.  The purpose of the dual encryption will become clear
in Section~\ref{sec:bait}.  We assume that the outer encryption is a fast
hybrid IND-CCA secure cipher such as Elliptic Curve El~Gamal with AES in OCB
mode~\cite{RogawayBB2003}.  We defer the specification of the inner
encryption scheme to Section~\ref{sec:crypto}.  It assures that, when the
decryptor receives the ciphertexts, it can verify the integrity of
individual chunks and of the message as a whole and he can associate the
chunks that belong to the same submission with all but negligible
probability (in $|k|$).



\subsection{Decryption}

It is substantially cheaper to multiply two DJ ciphertexts in the ciphertext
group than it is to decrypt one.  Furthermore, the product of ciphertexts
decrypts to the sum of the plaintexts in the plaintext group.  Recollect
that we expect to receive a large number of white ciphertexts, that is,
encryptions of zeroes.  This leads to the following optimization: we form a
full binary tree of fixed height, initialize its leaves with received
ciphertexts $c_1,\dots,c_n$ and initialize each inner node with the product
of its children.  Then, we begin to decrypt at the root.  If the plaintext
is zero then we are done with this tree, because all nodes in the tree are
zeroes.  Otherwise, the decryption yields $\gamma=\alpha+\beta\neq 0$ where
$\alpha,\beta$ are the plaintexts of the left and right child, respectively.
We decrypt the left child, which yields $\alpha$, and calculate the
plaintext of the right child as $\beta=\gamma-\alpha$ (without explicit
decryption).  If $\alpha$ or $\beta$ are zeroes then we ignore the
corresponding subtree.  Otherwise, we recurse into the subtrees that have
non-zero roots.  If a node is a leaf then we decrypt and verify it.  If we
find it invalid then we ignore the leaf.  Otherwise we forward its plaintext
to the file reassembly process.  We quantify the benefits of this algorithm
in Section~\ref{sec:treedec}.

\subsection{Software Dissemination}
\label{sec:getit}

We cannot simply offer the installer software for download because the
adversary would be able to observe that.  Instead, we pursue a multifaceted
approach to software distribution.
Our simplest and preferred approach involves the help of partners in the
print media business.  At the time of writing, popular print media often
come with attached CDROMs or DVDs that are loaded with, for example,
promotional material, games, films or video documentaries.  Our installer
software can be bundled with these media.
Our second approach is to encode the installer into a number of segments
using a Fountain Code.  In this approach, AdLeaks ads randomly request a
segment that the browser loads into the cache.  A small bootstrapper program
extracts the segments from the browser cache and decodes the installer from
it when enough of them have been obtained.  Since extraction happens outside
the browser it cannot be observed from within the browser.  The bootstrapper
can be distributed in the same fashion.  This reduces the distribution
problem to extracting a specific small file from the cache, for example, by
searching for a file with a specific signature or name in the cache
directory.  This task can probably be automated for most platforms with a
few lines of script code.  The code can be published periodically by trusted
media partners in print or verbatim in webpages or it could even be printed
on T-Shirts.
Our third approach is to enlist partners who bundle the bootstrapper with
distributions of popular software packages so that many users obtain it
along with their regular software.
With our multifaceted approach we hope to make our client software available
to most potential whistleblowers in a completely innocuous and unobservable
fashion.

\section{Ciphertext Aggregation}
\label{sec:crypto}

Our ciphertext aggregation scheme is based on the Damg\aa{}rd-Jurik (DJ)
scheme, which IND-CPA secure and is also an isomorphism of
\begin{align*}
  \psi_s &: \Za \times \Zm \leftrightarrow \Zn \\
  \psi_s(a;b) &\mapsto (1+N)^{a}\cdot b^{N^s} \imod N^{s+1}
\end{align*}
where $N$ is a suitable public key.  The parameter $s$ controls the ratio of
plaintext size and ciphertext size.  We use two two DJ encryptions $c,t$ to
which we jointly refer as a \emph{ciphertext.}  We refer to $t$ separately
as the \emph{tag.}  The motivation for this arrangement is improved
performance.  We wish to encrypt long plaintexts and the costs of
cryptographic operations increase quickly for growing $s$.  Therefore we
split the ciphertext into two components.  We use a shorter component with
$s=1$, which allows us to test quickly whether the ciphertext encrypts data
or a zero.  The actual data is encrypted with a longer component with $s>1$.
The two components are glued together using Pederson's commitment
scheme~\cite{Pederson1991}, which is computationally binding and perfectly
hiding.  This requires two additions to the public key, which are a
generator $g$ of the quadratic residues of $\Zm$ and some $h=g^x$ for a
secret $x$.  Instead of committing to a plaintext the sender commits to the
hash of the plaintext and some randomness.  We use a collision resistant
hash function $H$ for this purpose, which outputs bit strings of length
$|N/16|$.  Furthermore, let $R$ be a source of random bits.  The details of
the data encryption and decryption algorithms are as follows:
\begin{displaymath}
  \begin{array}[t]{l}
    \EncData(m, r_0) = \\
    \quad r_1, r_2 \leftarrow R \\
    \quad \text{chk} \leftarrow \text{if}~m,r_0=0 ~\text{then}~ 0 \\
    \qquad \text{else}~ H(m, r_0) \\
    \quad c \leftarrow \psi(m; h^{\text{chk}}\cdot g^{r_1})) \\
    \quad t \leftarrow \psi(r_0 || r_1;g^{r_2}) \\
    \quad \text{return} ~ c, t
  \end{array}
\end{displaymath}

\begin{displaymath}
  \begin{array}[t]{l}
    \DecVrfy(c,t) = \\
    \quad (m; k), (r_0 || r_1; \cdot) \leftarrow \psi^{-1}(c),
    \psi^{-1}(t) \\
    \quad \text{chk} \leftarrow \text{if}~m,r_0=0 ~\text{then}~ 0 \\
    \qquad \text{else}~ H(m, r_0) \\
    \quad \text{if}~ h^{\text{chk}}\cdot g^{r_1} = k ~\text{then} \\
    \qquad \text{return}~m, r_0\\
    \quad \text{return} ~ \bot
  \end{array}
\end{displaymath}
We assume that $|r_0|,|r_1|,|r_2|$ are polynomial in the security parameter.
Here, $r_0$ corresponds to $k||i||n$ as we introduced it in
Section~\ref{sec:prep}.  We define $\EncZero=\EncData(0,0)$.  Aggregation is
simply the multiplication of the respective ciphertext components.  We
establish the correctness of decryption next.  Let $c,t$ and $c',t'$ be two
ciphertexts.  Then
\begin{align*}
  c\cdot c'
  &=
  \psi(m+m';h^{H(m,r_0)+H(m',r_0')}\cdot g^{r_1+r_1'}) \\
  t\cdot t'
  &=
  \psi((r_0 || r_1)+(r_0' || r_1');g^{r_2+r_2'}) \\
  &=
  \psi((r_0+r_0') || (r_1+r_1');g^{r_2+r_2'})
\end{align*}
if $r_0,r_1,r_0',r_1'$ are left-padded with zeroes, which we hereby add to
the requirements.  The amount of padding determines how many ciphertexts we
can aggregate in this fashion before an additive field overflows into an
adjacent one and corrupts the ciphertext.  If we use $B$ bits of padding
then we can safely aggregate up to $2^B$ ciphertexts.  A length of $B=40$ is
enough for our purposes.  Observe that the aggregation of two outputs of
$\EncData$ is valid if and only if
\begin{align}\label{eq:H}
  H(m,r_0)+H(m',r_0') &\equiv H(m+m',r_0+r_0')
\end{align}
modulo $\phi(N)$.  This amounts to finding a collision in $H$ and we assume
that this is infeasible if $H$ resembles a pseudorandom function.  On the
other hand, if one input to the aggregation is an output of $\EncData$ and
another input is an output of $\EncZero$ then by the definitions of our
algorithms we have
\begin{align*}
  H(m,r_0)+0 &\equiv H(m+0,r_0+0)
\end{align*}
modulo $\phi(N)$, which is trivially fulfilled.  Hence, a valid data
encryption can be modified into another valid encryption of the same data
but not into a valid encryption of different data.  For this reason our
scheme is not IND-CCA secure, although it would achieve the weaker notion of
\emph{Replayable CCA}~\cite{CanettiKN2003} if we removed the special case
$m,r_0=0$.  Unfortunately we need \emph{some} special case to enable
aggregation.  Therefore we wrap the output of $\EncData$ into an IND-CCA
secure outer encryption in order to prevent adversaries from collecting
samples for use in a bait attack (see Section~\ref{sec:bait}).

\section{Security Properties}

\subsection{Traffic Analysis}

AdLeaks funnels all incoming transmissions to the decryptor, and
transmissions occur without any explicit user interaction.  Hence, the
posterior probability that anyone is a whistleblower, given his transmission
is observed anywhere in the AdLeaks system, equals his prior probability.
From that perspective, AdLeaks is immune against adversaries who have a
complete view of the network.  Furthermore, AdLeaks' deployment model is
suitable to leapfrog the long-drawn-out deployment phase of anonymity
systems that rely on explicit adoption.  For example, if \emph{Wikipedia}
deployed an AdLeaks script then AdLeaks would reach 10\% of the Internet
user population overnight, based on traffic statistics by
\emph{Alexa}~\cite{Alexa.com}.

\subsection{Denial of Service by Blocking}
\label{sec:br}  

AdLeaks is vulnerable to blocking.  Because anyone should be able to use
AdLeaks, anyone must be able to obtain the AdLeaks client software,
including the adversary.  It is easy to turn the client software into a
classifier that learns blocking filters for AdLeaks ads.  Furthermore, it is
harder in the AdLeaks case to bypass blocking than it is in the case of, for
example, censorship resistance software.  A whistleblower cannot trust
anyone and therefore it is very risky for him to obtain helpful information
by gossiping, which works for Tor.  Neither can he count on the help of
Internet Service Providers, which is the basis for systems such as
\emph{Telex}~\cite{WustrowWGH2011},
\emph{Cirripede}~\cite{HoumansadrNCB2011} and \emph{Decoy
  Routing}~\cite{KarlinEJJLMS2011}.  We are not aware of a practical
mechanism that is applicable to AdLeaks and that provides satisfactory
security guarantees.  Therefore, we defer countermeasure design to future
research.

\subsection{Denial of Service by Flooding}
\label{sec:dos} 

If we deployed one decryption unit and operated at its approximate limit,
that is, \Kmaxwb{} concurrent whistleblowers, then it would receive already
about \Kmaxfiles{} disclosures per day on average.  In other words, the
editorial backend of AdLeaks would be overwhelmed even before the technical
infrastructure is saturated.  Under these conditions, the benefit of
protecting against flooding attacks is debatable.

\subsection{Transmission Tagging}
\label{sec:tagging}

We have shown before that the encryption scheme AdLeaks uses is secure
against adaptive chosen ciphertext attacks as long as aggregators are
honest.  This prevents adversaries from tagging or adding chunks en route to
aggregators.  The Fountain code ensures that AdLeaks can determine when it
has enough chunks to recover the entire disclosure.  This prevents any
attempts to tag disclsoures by means of dropping or re-ordering chunks.

\subsection{Dishonest Aggregators}
\label{sec:bait}

Assume that AdLeaks did not use outer encryption.  Then adversaries might
employ the following \emph{active} strategy to gain information on who is
sending data to AdLeaks.  The adversary samples ciphertexts of suspects from
the network and aggregates the ciphertexts for each suspect.  He prepares a
genuine-looking disclosure that is enticing enough so that the AdLeaks
editors will want to publish it with high priority.  We call this disclosure
the \emph{bait.}  The adversary then aggregates suspects' ciphertexts to his
disclosure and submits it.  If AdLeaks does \emph{not} publish the bait
within a reasonable time interval then the adversary concludes that the
suspect is a whistleblower.  The reasoning is as follows.  If the suspect
ciphertexts were zeroes then the bait is received and likely published.
Since the bait was not published, the suspect ciphertexts carried data which
invalidated the bait ciphertexts.  This idea can be generalized to an
adaptive and equally effective non-adaptive attack that identifies a single
whistleblower in a group of $W$ suspects at the expense of $\log_2 W$ baits.
For this reason, AdLeaks employs an outer encryption which prevents this
attack.  However, if an adversary takes over an aggregator then he is again
able to launch this attack.  Therefore, aggregators should be checked
regularly, remote attestation should be employed to make sure that
aggregators boot the correct code, and keys should be rolled over regularly.
Note that it may take months before a disclosure is published and that a
convincing bait has a price --- the adversary must leak a sufficiently
attractive secret in order to make sure it is published.  From this, the
adversary only learns that a suspect has sent something but not what was
sent.

\subsection{Bandwidth-Based Attacks}

The adversary may submit a bait while sending a suspect chunk at a rate that
is close to or exceeds the data capacity between aggregators and decryptors.
If the suspect chunk is a zero then the recovery probability of AdLeaks
remains within the expected bounds.  If, however, the suspect chunk is a
data chunk then, with good probability, AdLeaks does not recover the bait.
However, AdLeaks will notice the reduced recovery probability and may react,
for example, by rolling over to a new key or even by pushing warnings to
whistleblowers via its ad distribution mechanism.

\subsection{Client Compromise}

If the computer of a whistleblower is compromised then the whistleblower
loses all security guarantees.  Limited resilience to detection could
perhaps be achieved by employing malware-like hiding-tactics.  However,
since the AdLeaks client is public, it is merely a matter of time until its
tactics are reverse engineered and a detection software is written.  If the
detection software can be pushed to suspects' computers then whistleblowers
can be uncovered.  In order to limit the risk, a ``production-grade''
implementation of the AdLeaks client should offer a function to delete
itself securely when it is not needed anymore.

\section{Scalability}

Our goal in this section is to characterize how well AdLeaks scales with a
growing number of users and whistleblowers.  Among other dimensions, we
explore the necessary and sufficient size of the required infrastructure and
the time it takes to submit a file.  An estimate of the financial
feasibility of the operation can be found in Section~\ref{sec:financial}.

\subsection{Submission Duration}

In the absence of better data, we analyzed the \emph{Wikileaks} archives
available from \emph{wlstorage.net} and estimated the sizes of disclosures
as follows: we counted top-level files and archives as individual
disclosures; we counted the contents of subdirectories as a single
disclosure unless the contents also appeared in an archive.  We found that
$70$\% of the disclosure estimates were less than 2~MB (about $20$\% were
larger than $4$~MB) and chose $\Kfile$~\Ufile{} to be our target size.

Zhang and Zhao conducted a study on tabbed browsing
behavior~\cite{ZhangZ2011} and measured 89,851 page loadings distributed
over~20 participants and~31 days, which amounts to about~145 page loadings
per user per day.  Webpages often display multiple ads.  It is common to
display a horizontal ad at the top and one or more vertical ads in the
margins.  The popularity of a website also plays an important role for how
many ad loadings it can trigger.  News websites elicit frequent and regular
page loadings and are particularly suitable for our purposes.  Fortunately
for us, they also have incentives to support online whistleblowing systems.
Furthermore, they might support persistent ``ad-less'' ads, that is, AdLeaks
scripts that do not take up page real estate and therefore do not compete
with other ad revenue sources for the news outlet.  Lastly, our ads can
trigger multiple transmissions at random intervals if we are below our
target.  For this reason, we decided that it is fair to assume we would get
about $\Kads$ transmissions per day and user from our ads.

For a sound level of security, we use a public key modulus with 2048 bits.
The DJ scheme is expensive for increasing plaintext lengths and based on
initial tests we settled on parameters that support $\Kmlen$ \Umlen{} for
data use.  Our file would thus require about $\Kcblk$ blocks.  If we account
for the encoding overhead and further assume that AdLeaks can recover a data
transmission with probability at least $\Kp$ then submitting a
$\Kfile$~\Ufile{} file requires about $\Kcnum$ transmissions or $\Kdays\pm1$
days on average.

\subsection{Network Load}

A Base64 encoded ciphertext requires transmitting about \Ktlen{} bytes if we
include \Kheaders{} bytes worth of HTTP headers.  At \Kads{} transmissions a
day this adds up to an additional daily network load of
\Kloadinc~\Uloadinc{}.  Since flat rates for Internet access are common in
developed countries, we believe this is insignificant for users in these
countries.  Also, given that whistleblowers provide large societal benefits,
society may well be willing to pay even a small, but significant, cost in
bandwidth, beyond that needed by AdLeaks.
Although one might get concerned that the accumulated load poses a problem
at Internet scale.  This is not likely the case.  The quoted amount of data
is less than the size of an average web page on the wire~\cite{Google2010},
which Google engineers found to be about 320KB in 2010.  Hence, the load
AdLeaks adds to the network is well within users' network traffic variance.
It may not even be noticed against the backdrop of video streaming and
increasing Internet use.

\subsection{Guards and Aggregators}
\label{sec:lab}

In order to establish a lower bound of the request rate that AdLeaks servers
would be able to process we deployed guard and aggregator prototypes (see
Section~\ref{sec:impl}) on EmuLab~\cite{WhiteLSRGNHBJ2002}.  We used six
pc3000 nodes (3.0~GHz 64-bit Xeon processors, 2~GB DDR2 RAM, 1~Gbit
connectivity) for guards, one d710 node (2.4~GHz 64-bit Quad Core Xeon E5530
processor, 12~GB RAM, 1~Gbit connectivity) for one aggregator and another
pc3000 node for one decryptor.  The guard servers sent chunks to the
aggregator through a reverse SSH tunnel.  The aggregator aggregated the
incoming chunks and sent aggregates to the decryptor through a reverse SSH
tunnel.  The decryptor performed the tree decryption and discarded the
decrypted chunks.  Overload situations were easily observed through buffer
growth, that is, the incoming rate exceeded the rate at which the aggregator
processed and sent its aggregates.  The aggregator operated stable at an
incoming rate of \Kmaxreqii{} chunks per second or roughly \Kmbsii~\Umbsii.
Since our prototype does not yet implement the outer encryption we measured
the overhead of elliptic curve based key exchanges separately.  For a
\Keccklen{}-bit key we measured \Keccmean~\Ueccmean{} with a standard
deviation of \Keccsdev~\Ueccsdev{} (100 iterations).  If we correct for this
overhead then the aggregator handles \Keccreqii~\Ueccreqii.  Since guards
merely strip some encoding from incoming requests and pass them on we did
not measure them separately and instead assume that they perform similar to
aggregators without outer encryption, that is, we assume they handle
\Kmaxreqi~\Umaxreqi{}.

Web surfers are most active during certain time windows during the day.  If
active times are reasonably uniform across the population then they shift
with the time zone, which spreads the active windows.  This prompts the
following assumptions, which may capture, for example, the situation in the
United States: the active time of users is between 6pm and 1am, and they
live in three adjacent time zones.  This means, conservatively, that the
$\Kads$ transmissions per user we assumed before concentrate within an
$\Khours$ hours window. The U.S.\@ have about \Kusers{} \Uusers{} broadband
Internet users at the age of 18 years and older~\cite{Stats2012a}.
The expected load on guards is therefore about $\Kreqi$ \Ureqi{}.  However,
we are rather interested in the \emph{peak} load, for example, the load that
is not exceeded in 99\% of all cases.  Towards this end we model arrival
times using a Poisson distribution.  For the calculation of the peak load we
use the fact that the Poisson distribution is very close to the normal
distribution for a mean larger than about~20.  It is well known that about
99.7\% of normally distributed events are within 3 standard deviations from
the mean.  Since the variance of the Poisson distribution equals its mean we
have that, in 99.85\% of all cases, the load will be at most \Ksreqi{}
\Usreqi.  Using this as our basis we conclude that we need \Kunitsi{} guard
units and \Kunitsii{} aggregators.

\subsection{Data Recovery}
\label{sec:recp}
\label{sec:treedec}

By design, the decryptor scalability does not depend on the number of users
but only on the number of whistleblowers.  This is what allows us to upper
bound the number of copies of the AdLeaks decryption key.  Based on speed
test reports on regional ISPs their download bandwidths range from 4 megabit
per second (Mb/s) to 31 Mb/s for a household Internet connection.  As a
basis for subsequent estimation we use the median rounded to Mb, that is, we
assume that decryptors can download ciphertexts from aggregators with
$\Kcapiii$~\Ucapiii.  At $\Kclen$ \Uclen{} per ciphertext this translates to
$\Kpktiii$ aggregate ciphertexts per second.

\begin{figure}
  \includegraphics[width=\columnwidth]{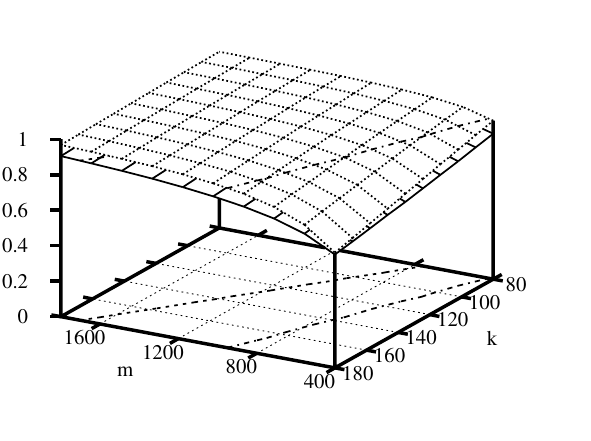}
  \caption{Shows the graphs of the recovery probability for varying numbers
    of data chunks and aggregates for $t=1$ and $t=4$.  Contour lines
    indicate a $0.9$ probability level.}
  \label{fig:recovery3d}
\end{figure}

We estimate the data recovery probability of AdLeaks under the following
aggregation model.  Given $k$ gray ciphertexts and an arbitrary number of
white ones, what is the probability that the data from a random gray
ciphertext can be recovered if we can transmit $m$ aggregates from
aggregators to decryptors?  The data of a gray ciphertext is recoverable if
it is aggregated only with white ones.  Therefore, we seek the probability
that, given any gray ciphertext, all other gray ciphertexts fail to be
assigned to the same aggregate, of which there are $m$.
If $k$ is small with respect to $m/t$ for some $t$ then we can improve our
recovery probability as follows.  We perform $t$ independent aggregations
with $t$ sets of $m/t$ aggregates.  This yields the same overall number $m$
of aggregates.  A gray ciphertext is recoverable if it is recoverable from
any of the $t$ sets.  Hence, the probability we seek is one minus the
probability that we fail to recover the data from all of the $t$ sets of
aggregates.  Both probabilities are given by the following formulas:
\begin{align*}
  \Pr[i=1] &= (1-\frac{1}{m})^{k-1}  \\
  \Pr[i=1] &= 1-(1-(1-\frac{1}{m/t})^{k-1})^t
\end{align*}
Figure~\ref{fig:recovery3d} illustrates the effect for $t=1$ and for $t=2$.
The contour lines indicate where the recovery probability becomes larger
than $\Kp$.
For example, if aggregators can transfer $\Kpktiii$ aggregates per second to
a decryptor and whistleblowers send at most $\Kmaxdata$ gray ciphertexts per
second and we set $t=1$ then AdLeaks can recover each transmission with a
probability of at least $\Kp$.

\begin{figure}
  \includegraphics[width=\columnwidth]{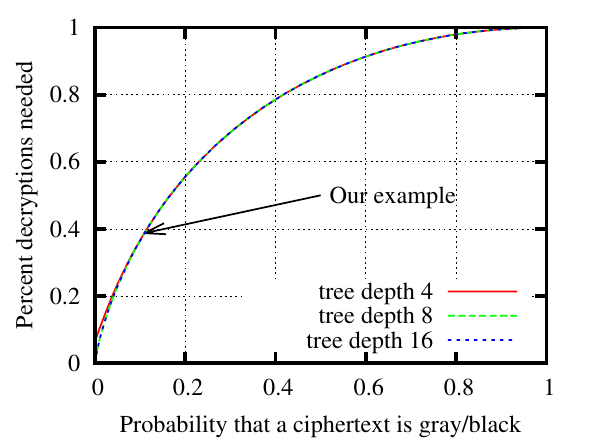}
  \caption{Shows the normalized savings of the tree decryption algorithm for
    various tree sizes and load characteristics.  For $t>8$ the graph looks
    identical to the graph for $t=8$.}
  \label{fig:cryptcost}
\end{figure}

Next, we estimate the number of cryptographic operations AdLeaks must
perform in order to decrypt with its tree decryption algorithm.  Assume
AdLeaks builds trees of depth $n$.  This requires $2^{n}-1$ modular
multiplications.  The expected number of decryptions is
\begin{align*}
  E(X) &= 1 + \frac{1}{2}\cdot \sum_{i=1}^{n} 2^{i} \cdot
  (1-(1-p)^{2^{n+1-i}})
\end{align*}
\begin{proof}
  The root of the tree must always be decrypted, hence the expectation is
  always at least $1$.  Since the algorithm only decrypts left children and
  not right children, we need to count only half of the remaining nodes.
  Recall that a right child is calculated by subtracting its sibling from
  its parent.  At level $i$ from the root, starting under the root node, we
  have $2^i$ nodes.  We have to decrypt a left child at level $i$ if its
  parent is not zero.  The probability that the parent is not zero is one
  minus the probability that its $2^{t+1-i}$ leaves are all zeroes.
\end{proof}
If we divide the expected number of decryptions by $2^n$ decryptions (the
na\"\i{}ve approach) and plot the normalized results for several values of
$n$ then we obtain the graphs in Figure~\ref{fig:cryptcost}.  The graphs tell
us, for example, that we expect to save $\Ksavdec\%$ of the decryptions if
AdLeaks operates at its limit, that is, a recovery probability of $\Kp$ and
$\Kmaxdata/\Kpktiii \approx \Kpdata\%$ gray or black ciphertexts.  The
lighter the load is the more we save.

\subsection{Number of Whistleblowers}
\label{sec:numwb}

We characterize next how many concurrent whistleblowers a link capacity of
$\Kmaxdata$ data chunks per second can support.  Since no data set exists
that we could use to estimate the arrival times of data chunks, we model
them as a Poisson process that we approximate by a normally distributed
process as before.  To be on the safe side we seek a safe average sending
rate $r$ so that the actual sending rate does not exceed $\Kmaxdata$
\Umaxdata{} in 97.725\% of all cases, that is, the second quantile.  The
safe average rate can be found easily by solving $r+2\cdot\sqrt r=\Kmaxdata$
for $r$, which yields \Kavgdata{} \Uavgdata.  At \Kads{} transmissions per
day and whistleblower this means that AdLeaks can serve \Kmaxwb{} concurrent
whistleblowers at any time with a \Kcapiii{} \Ucapiii{} uplink for the
decryptor.

\subsection{Decryptors}

The performance of the decryptor is bound by the cost of two operations: the
time it takes to test whether a tree node is white, and the time it takes to
perform a full decryption and verification.  We measured \Ktestmean{} and
\Kdecmean{} seconds for these operations on a single core (2.66GHz Intel
Xeon) with negligible standard deviation.  If we assume that the decryptor
has \Kcores{} cores available for decryption then we estimate that our
running example requires \Kunitsiii{} decryptors.  Since the necessary
resources for decryption scale linearly with the number of whistleblowers
this means that AdLeaks can serve about $\Kunitwb$ concurrent whistleblowers
with just one unit, for example, a dual 6-core Mac Pro.

\subsection{Client Measurements}

Our JavaScript implementations of the DJ scheme leverage several
optimizations~\cite{Jurik2003} that improve efficiency over unoptimized DJ
by a factor of 8 to 32 in our measurements. As a side effect, the bit
lengths of two parameters of our inner encryption scheme, namely $r_1,r_2$
(see Section~\ref{sec:crypto}), bound the time it takes to encrypt in the
browser.  Reasonable values range from 512 bits (probably sufficient) to
2044 bits (paranoid).  We measured these times on an Intel i5-2500K CPU at
3.30~GHz with Chromium Version 20.0.1132.47.  For 512 bits it took
\Kodjlmean{} seconds ($s=\Kodjlsdev$, speedup $\approx 32$), for 1024 bits
it took \Kodjmmean{} seconds ($s=\Kodjmsdev$, speedup $\approx 16$) and for
2044 bits it took \Kodjhmean{} seconds ($s=\Kodjhsdev$, speedup $\approx
8$).  Since Java is still significantly faster than JavaScript we assume
that these times will become smaller still in the future.

\section{Financial Viability}
\label{sec:financial}

Given the server resources AdLeaks requires it is prudent to ask what are
the costs the AdLeaks organization has to bear.  We found that dual
quad-core servers with unmetered 1Gb interfaces are available for less then
\Kunitmonth~\Uunitmonth.  At this price, the infrastructure for the guards
and aggregators necessary to serve \Kusers~\Uusers{} users would cost
\Kdailyusd~\Udailyusd{}.  While this seems high it is instructive to look at
the revenue side.  Each ad loading corresponds to what is called an
``impression'' in the online advertising business.  The prices for
impressions are typically quoted as \emph{costs per mille,} or CPM.  Top
ranking websites can command prices over 100~USD while run-off-the-mill
websites receive in the order of \Kcpm~\Ucpm.  \emph{Cost per click} models
or \emph{cost per conversion} models generate additional revenue, which we
ignore for the sake of conservatism.  If AdLeaks had \Kusers~\Uusers{} users
who see \Klowads{} AdLeaks ads per day on average and if AdLeaks paid out
\Kcpm~\Ucpm{} per mille impressions and if its markup was \Kbreakeven{}
\Ubreakeven{} or better then AdLeaks would break even on its infrastructure
costs.  For comparison, ValueClick Inc.\@ reported a gross profit of over 96
million USD on a revenue of about 161 million USD in its second quarter of
2012, which translates to about 59 percent profitability, and reported 25
cents of net income per common share.  The numbers suggest that, if AdLeaks
was run as a not-for-profit operation, it could gain market share by
offering very competitive pricing while earning a decent plus.  This is in
sharp contrast to contemporary whistleblowing platforms who depend entirely
on donations.

\section{Implementation}
\label{sec:impl}

We developed fully-functional multi-threaded aggregation and decryption
servers with tree decryption support as well as a Fountain Code encoder and
decoder.  Decryptors write recovered data to disk and the decoder recovers
the original file.  We also developed a fake guard server which is capable
of generating and sending chunks according to a configurable ratio of white
and gray ciphertexts.  All servers connect to each other through SSH tunnels
via port forwarding.  The entire implementation consists of 101 C, header
and CMake files with 7493 lines of code overall.  This includes our
optimized DJ implementation~\cite{Jurik2003}, which is based on a library by
Andreas Steffen, a SHA-256 implementation by Olivier Gay, and several
benchmarking tools.  Our ads implement the DJ scheme based on the
\emph{JSBN.js} library and use \emph{Web Workers} to isolate the code from
the rest of the browser.  The entire ad currently measures less than 81~KB.
The size can be reduced further by eliminating unused library code and by
compressing it.  The ad submits ciphertexts via \emph{XmlHttpRequests.}  We
instrumented the Firefox browser for our prototype and patched the source
code in two locations.  First, we hook the compilation of \emph{Web Worker}
scripts and tag every script as an AdLeaks script if it is labeled as one in
lieu of carrying a valid signature.  We placed a second hook where Firefox
implements the \emph{XmlHttpRequest.}  Whenever the calling script is an
AdLeaks script running within a \emph{Web Worker,} we replace the zero chunk
in its request with a data chunk.

\section{Related Work}


Closely related to our work, there is early work on DC Nets
\cite{Chaum1988,WaidnerP1989} which aims to provide a cryptographic means to
hide who sends messages, the use of Raptor codes \cite{CastiglioneDFP2012}
to implement an asynchronous unidirectional one-to-one and one-to-many
covert channel using spam messages, and anonymous data
aggregation~\cite{PuttaswamyBP2010} for distributed sensing and
diagnostics. In preserving the privacy of web-based
email~\cite{ButlerEPTM2006} one can take advantage of a spread-spectrum
approach for hiding the existence of a message, but it is not secure against
a global attacker. Membership-concealment~\cite{VassermanJTHY2009} can also
be used to hide the real-world identities of participants in an overlay
network, but this doesn't suffice for AdLeaks.

In censorship resistance, there is Publius~\cite{WaldmanRC2000} which is an
anonymous publication system but does not offer any sort of connection-based
anonymity. Collage~\cite{BurnettFV2010} steganographically embeds content in
cover traffic such as photo-sharing sites and implements a rendezvous
mechanism to allow parties to publish and retrieve messages in this cover
traffic, but Alice and Bob must exchange a key a priori. More recent
work~\cite{InvernizziKV2012} explores an approach that assumes the ability
of being able to globally check and retrieve all blog posts in real time and
determining and extracting all the embedded content.

Another related area is secure data aggregation in wireless sensor
networks~\cite{Alzaid2011} or WSNs. One can try to securely aggregate
encrypted data~\cite{Castelluccia2009}, which identifies the key stream in
the header and requires removing a stream for each ciphertext received. The
latter wouldn't scale for AdLeaks because it requires millions of keystream
removals per aggregate. One approach~\cite{ViejoWD2012} for aggregating in
multicast communication uses the Okamoto-Uchiyama encryption scheme for
secure aggregation, which resembles Pederson's commitment scheme very
closely. The difference is really that AdLeaks is not used in the aggregate
but instead deals with collisions. Other work~\cite{Wagner2004,ChanPPS2007}
targets the security of statistical computations on the inputs from various
sensor nodes. The two key differences in the approaches found in WSNs are:
(i) WSNs want correctly aggregated data that allows for unencrypted sending,
whereas our approach seeks the opposite of that, and (ii) the attacker wants
to have a tainted input accepted, whereas in our context he wants to learn
the content of the input. In WSNs both event-driven and query-based
processing is of interest, with most approaches focusing on query-based
solutions, whereas AdLeaks is event-driven. That also means that we don't
know \emph{a priori} which client sends what in what round. It is difficult
for us to exchange keys beforehand and our approach remains unidirectional,
i.e. we cannot distribute keys. Lastly, in WSNs clients are not trusted
initially and are later vetted, whereas in our approach clients are never
trusted.

\section{Conclusions}

AdLeaks leverages the ubiquity of online advertising to provide anonymity
and unobservability to whistleblowers making a disclosure online. The system
introduces a large amount of cover traffic in which to hide whistleblower
submissions, and aggregation protocols that enable the system to manage the
huge amount of traffic involved, enabling a small number of trusted nodes
with access to the decryption keys to recover whistleblowers' submissions
with high probability. We analyzed the performance characteristics of our
system extensively. Our research prototype demonstrates the feasibility of
such a system. We expect many aspects of the system can be improved and
optimized, providing ample opportunity for further research.

\subsection*{Acknowledgements}

The first, second and last author are supported by an endowment of
\emph{Bundesdruckerei GmbH.}  An abridged version of this paper has been
accepted for publication in the proceedings of \emph{Financial Cryptography
  and Data Security 2013}~\cite{RothGRDR2013}.  Copies of the abridged
version will eventually become available at
\url{http://www.springer.de/comp/lncs/}.

\flushleft

\end{document}